\documentclass[12pt]{article}
\usepackage{amsmath,amssymb,epsfig,amsfonts}
\usepackage{graphicx,subfigure}
\usepackage[usenames, dvipsnames]{color}
\usepackage[backref]{hyperref}
\usepackage{cite}
\makeatletter

\newcommand{\be}{\begin{equation}}
\newcommand{\ee}{\end{equation}}
\newcommand{\ba}{\begin{aligned}}
\newcommand{\ea}{\end{aligned}}

\def\unit{{1\kern-.65ex {\rm l}}}
\def\1{{1\kern-.65ex {\rm l}}}

\newcount\hour \newcount\minute
\hour=\time \divide \hour by 60
\minute=\time
\def\now{%
\ifnum \hour<13
  \ifnum \hour=0 \advance \hour by 12 \number\hour:\else \number\hour:\fi%
     \ifnum \minute<10 0\fi%
     \number\minute%
\ A.M.%
\else \advance \hour by -12 \number\hour:%
  \ifnum \minute<10 0\fi%
  \number\minute%
  \ P.M.%
\fi%
}

\makeatother

\begin{document}

% format
\baselineskip=18pt  % a la harvmac
\numberwithin{equation}{section}  % make eq labels (sec.num)
\allowdisplaybreaks  % allow page breaks in displayed eqs

% print date, time and filename 
%\pagestyle{myheadings}
%\markright{{\tt \jobname.tex} -- \today{} \now}

%%%%%%%%%%%%%%%%%%%%%%%%%%%%%%%%%%%%%%%%%%%
%%%        TITLE BEGINS HERE
%%%%%%%%%%%%%%%%%%%%%%%%%%%%%%%%%%%%%%%%%%%

%% ========== title (note version) begins here ==========
%
%\vspace*{-1cm}
%\begin{center}
% {\Large\bf Title of the Document}
%\end{center}
%\vspace*{-.5cm}
%
%% ========== title (note version) ends here ==========

%% ========== title (paper version, a la harvmac) begins here ==========

\thispagestyle{empty}

\vspace*{-2cm} 
\begin{flushright}
{\tt LTH-949}\\
\end{flushright}

% title, authors, affiliation
\vspace*{0.8cm} 
\begin{center}
 {\LARGE GUT theories from Calabi-Yau 4-folds with $SO(10)$ Singularities \\}
 \vspace*{1.5cm}
 Radu Tatar  and William Walters \\
{\it   Department of Mathematical Sciences, University of Liverpool\\
Liverpool, L69 7ZL, England\\}
{\tt radu.tatar at liverpool.ac.uk , wrw23 at liverpool.ac.uk}\\
\vspace*{0.8cm}

\vspace*{0.8cm} 
% {\tt foo@bar}, 
% {\tt sh{}i{}ge{}@t{}h{{}}{e{ory}}.c{}a{}l{{}}te{{}}ch.e{{}}d{}u} % can this avoid spam?
\end{center}
\vspace*{.5cm}

% abstract
%\noindent
%
%Insert Abstract Here

We consider an $SO(10)$ GUT model from F-theory compactified on an elliptically fibered Calabi-Yau with a $D_5$ singularity. 
To obtain the matter curves and the Yukawa couplings, we use a 
global description to resolve the singularity along the lines of \cite{sm1}.
We identify the vector and spinor matter representations and their Yukawa couplings and we explicitly build the 
G-fluxes in the global model and check the agreement with the semi-local results. As our bundle is of type $SU(2 k)$, 
some extra conditions need to be applied to match the fluxes, as observed in \cite{sak}.   

\newpage
\setcounter{page}{1} % don't number title page

%% ========== title (paper version, a la harvmac) ends here ==========

%%%%%%%%%%%%%%%%%%%%%%%%%%%%%%%%%%%%%%%%%%%
%%%           TITLE ENDS HERE
%%%%%%%%%%%%%%%%%%%%%%%%%%%%%%%%%%%%%%%%%%%

\tableofcontents
%\printindex

%%%%%%%%%%%%%%%%%%%%%%%%%%%%%%%%%%%%%%%%%%%
%%%        MAIN TEXT BEGINS HERE
%%%%%%%%%%%%%%%%%%%%%%%%%%%%%%%%%%%%%%%%%%%
\newpage 
\section{Introduction}
In recent years, F-theory phenomenology has been one research avenue providing insights into Beyond Standard Model physics. One of the successes of F-theory compactification has been the fact that it allows an interpretation for the Yukawa couplings
in GUT theories. Firstly discussed in terms of heterotic-F theory duality \cite{tw1}, the introduction of local F-theory models as 8-dimensional field theories\cite{bhv1,dw1} allowed some detailed 
studies of the localisation of Yukawa couplings on Calabi-Yau 4-folds, one of them being developed in \cite{tw2,tw3,tw4} 
(for other methods and a full set of references see \cite{weigand}). To study the Yukawa couplings, one is tempted to follow the procedure involving the matter fields which were described some time ago in \cite{kv}. When considering the Yukawa couplings in an $SU(5)$ GUT, a nice feature of the
F-theory description is that the up-type Yukawa couplings appear at enhancements of $A_4$ singularity to $E_6$ singularity in 
codimension-three, and there is no perturbative type IIB picture for this enhancement. The singular fibre is expected to degenerate from $A_4$ in codimension 1 to either $A_5$ or $D_5$ in codimension 2 and to either $A_6$, $D_6$ or $E_6$ in codimension 3.   

When considering the heterotic - F theory duality \cite{tw2, tw3}, the heterotic results were mapped into the local F-theory as a choice of a diagonal 
expectation value for an 8-dimensional Higgs field (see also \cite{wa}) \footnote{If the Higgs field is not diagonal, one could use T-branes which are non-abelian bound states of branes \cite{cchv,cftw}}. The non-zero values for the vevs of the Higgs field describe the deformation $E_6 \rightarrow A_4$ after taking a singular limit of $A_4$ and mapping in into the $E_6$ \cite{tw3}. Nevertheless, when applied to the global geometry, this involves applying Tate's algorithm for codimension 3 loci which is outside its validity. 

A detailed study of the F-theory geometries was started in \cite{ey} (see also \cite{mt}) for $SU(5)$ GUT theories by considering an explicit blow-up of $A_4$ singularities. For a general choice of complex structure, the singularity is resolved with four additional $P^1$ cycles. When some complex deformations are turned off, one encounters some lines of singularities and some additional $P^1$ cycles are needed to obtain a smooth manifold.
The intersection of the full set of resolution cycles is supposed to reproduce the Dynkin diagram  for $D_5, A_5$ for matter or $D_6, E_6, A_6$ for Yukawa couplings. When studying the intersections of the full set of cycles, \cite{ey} observed that they do not fully reproduce the Dynkin diagrams for the higher singularities, the biggest departure being for the expected $E_6$ singularity which would provide the up-type Yukawa couplings. 

The discussion of the local geometry in \cite{ey} was reconsidered in \cite{sm1} by looking at global geometries and utilising their previous developed formalism 
\cite{sm2,sm3,sm4} (see also \cite{wei,gh} for similar approaches and \cite{sak} for using the same method in the case of the $E_6$ model). In the global formalism, the divisors correspond to zeroes of various sections and the cycles associated to the $A_4$ roots appear as intersections between various divisors. At singular loci the number of irreducible components increases and some of the roots split due to the weights of $\mathbf{5}$ or $\mathbf{10}$ representations. One important aspect of the global geometries is that the number of roots does not increase and the rank of the singularity does not enhance.

In the current work we consider an $SO(10)$ GUT model which corresponds to a $D_5$ singularity. $SO(10)$ models were previously considered within the F-theory context in \cite{so101,so102,so103} and we focus on the global construction, by resolving the singularity and studying the points in the complex structure where extra blow-ups are needed. In section 2 we construct the resolution of a singularity of type $D_5$. In section 3.1 we describe the matter in the spinor 16 representation, in section 3.2. we consider matter in the vector 10 representation and in section 3.3 we discuss the Yukawa couplings. In section 4 we compute the local and the global fluxes. The matching is obtained in the particular case when the first Chern class of the base or the  first Chern class of the divisor are even.

\section{Resolution: Generalities}
\label{sec:generalities}

\subsection{Setup}

We consider an $SO(10)$ GUT group in F-theory phenomenology which corresponds to compactifying on a Calabi-Yau 4-fold obtained after 5 resolution steps applied to a $D_5$ singularity. In this section we are going to explicitly describe the procedure to obtain the smooth Calabi-Yau 4-fold. 

As in the case of $SU(5)$, we construct the resolution in the auxiliary 5-fold
\begin{equation}X_5=\mathbb{P}\left({\cal{O}}\oplus K_{B_3}^{-2}\oplus K_{B_3}^{-3}\right) \,.
\end{equation}
$X_5$ is a $\mathbb{P}^2$ bundle, the divisors on $X_5$ consist of pullbacks of divisors on $B_3$ under the projection
\begin{equation}\pi_X:X_5\rightarrow B_3
\end{equation}
together with a new divisor, $\sigma$, inherited from the hyperplane of the $\mathbb{P}^2$ fiber. 
The projective coordinates $w$, $x$, and $y$ on the $\mathbb{P}^2$ fiber of $X_5$ are sections of the following bundles on $X_5$
\begin{equation}\begin{array}{c|c}
\text{Section} & \text{Bundle} \\ \hline
w & {\cal{O}}(\sigma) \\
x & {\cal{O}}(\sigma+2c_1) \\
y & {\cal{O}}(\sigma+3c_1)\\
z & {\cal{O}}(S_2) \\
\end{array}\label{wxysections}\end{equation}
$z$ is the section that vanishes along $S_2$.
The  Tate form  for an $SO(10)$ singularity at $z=0$ is 
\begin{equation}\label{TateSO10}
y^2 w + b_1 z x y w + 
 b_3 z^2 y w^2  =  x^3 + b_2 z x^2 w + b_4 z^3 x w^2 + b_6 z^5 w^3 \,. 
  \end{equation}

%%%%%%%%%%%%
%%%%%%%%%%%%
 
\subsection{Resolution of $D_5$ singularity}

We now embark in the procedure to resolve the $D_5$ singularity, involving five blowups.  

%%%%%%%%%%%%

\subsubsection{First Two Blowups}

The locus 
\be
x=y=z=0
\ee
is singular. To blow up along it, we introduce 
\begin{equation}x = \zeta x_1\,,\qquad y = \zeta y_1\,,\qquad z = \zeta z_1\,,
\end{equation}
where $\zeta=0$ gives rise to an exceptional divisor $E_1$. 
The new classes of the sections are 
\begin{equation}\begin{array}{c|c}\text{Section} & \text{Bundle} \\ \hline
x_1 & {\cal{O}}(\sigma+2c_1 - E_1) \\
y_1 & {\cal{O}}(\sigma+3c_1 - E_1) \\
z_1 & {\cal{O}}(S_2-E_1) \\
\zeta & {\cal{O}}(E_1)
\end{array}\end{equation}
After a proper transform, the equation for $Y_4$ becomes
\be
w \left(-\zeta  z_1 \left(\zeta  w z_1^2 \left(b_6 \zeta  w^3 z_1^2+b_4 x_1\right)+b_2
   x_1^2\right)+\zeta  y_1 z_1 \left(b_3 w z_1+b_1 x_1\right)\right)+w y_1^2-\zeta  x_1^3=0
\ee

The second blowup is along $x_1=y_1=\zeta=0$, which is obtained by setting
\begin{equation}
x_1 = x_2\alpha \,,\qquad y_1 = y_2\alpha \,,\qquad \zeta = \zeta_2\alpha \,.
\end{equation}
The section $\alpha=0$ gives rise to an exceptional divisor $E_2$. 
The new sections are
\begin{equation}\begin{array}{c|c}\text{Section} & \text{Bundle} \\ \hline
x_2& {\cal{O}}(\sigma+2c_1-E_1-E_2) \\
y_2 & {\cal{O}}(\sigma+3c_1-E_1-E_2) \\
\zeta_2 & {\cal{O}}(E_1-E_2) \\
\alpha & {\cal{O}}(E_2)
\end{array}\end{equation}
The proper transform of the fourfold defining equation is 
\be
w y_2 \left(\zeta _2 z_1 \left(b_3 w z_1+\alpha  b_1 x_2\right)+y_2\right)
= \alpha  \zeta _2
   \left(b_6 \zeta _2^2 w^3 z_1^5+b_4 \zeta _2 w^2 x_2 z_1^3+b_2 w x_2^2 z_1+\alpha  x_2^3\right)
\ee

%%%%%%%%%%%%%%%%%%

\subsubsection{Third Blowup}
We can then blow up along $y_2=\zeta _2=\alpha=0$, which we do by setting
\begin{equation}
y_2 = y_3\beta \,,\qquad \zeta _2 = \zeta _3\beta \,,\qquad \alpha = \alpha _3\beta \,.
\end{equation}
The section $\beta=0$ gives rise to a new exceptional divisor $E_3$.
So the new sections are
\begin{equation}\begin{array}{c|c}\text{Section} & \text{Bundle} \\ \hline
y_3& {\cal{O}}(\sigma+3c_1-E_1-E_2-E_3) \\
\zeta _3 & {\cal{O}}(E_1-E_2-E_3) \\
\alpha _3 & {\cal{O}}(E_2-E_3) \\
\beta & {\cal{O}}(E_3)
\end{array}\end{equation}
The proper transform of the equation for $Y_4$ is
\begin{equation}
w y_3 \left(\zeta _3 z_1 \left(b_3 w z_1 + \alpha _3 \beta b_1 x_2 \right) +y_3 \right)
= \alpha _3 \zeta _3  \left( b_6 \beta ^2 \zeta _3^2 w^3 z_1^5+b_4 \beta \zeta _3 w^2 x_2 z_1^3
+b_2 w x_2^2 z_1+\alpha _3 \beta x_2^3\right)
\end{equation}

%%%%%%%%%%%%%%%%%%%%

\subsubsection{Final Two Blowups}

The last two blowups are carried out as in \cite{saklaw}. The fourth blow up is along $y_3=\zeta_3=0$ and we do this by setting
\be 
y_3=y_4 \delta_4 \,, \qquad \zeta_3=\zeta_4 \delta_4 \, .
\ee 
The proper transform is
\be 
\ba 
wy_4^2 \delta_4 +b_1 wx_2y_4 z_1 \zeta_4 \alpha_3 \beta \delta_4  + b_3 w^2 y_4z_1 \zeta_4 \delta_4 =\,& x_2^3 \zeta_4 \alpha_3^2 \beta + b_2 wx_2^2 z_1 \zeta_4 \alpha_3 +b_4 w^2 x_2 z_1^3 \zeta_4^2 \alpha_3 \beta \delta_4 \cr
& +b_6 w^3 z_1^5 \zeta_4^3 \alpha_3 \beta^2 \delta_4^2 
\ea 
\ee

The fifth blow up is along $y_4=\alpha_3=0$ and is given by
\be 
y_4=y_5 \delta_5 \,, \qquad \alpha_3=\alpha_5 \delta_5 \, ,
\ee
giving a proper transform
\be 
\ba 
wy_5^2 \delta_4 \delta_5+b_1 wx_2y_5 z_1 \zeta_4 \alpha_5 \beta \delta_4  \delta_5 + b_3 w^2 y_5z_1 \zeta_4 \delta_4 =\,& x_2^3 \zeta_4 \alpha_5^2 \beta \delta_5 + b_2 wx_2^2 z_1 \zeta_4 \alpha_5  \cr
&+b_4 w^2 x_2 z_1^3 \zeta_4^2 \alpha_5 \beta \delta_4 +b_6 w^3 z_1^5 \zeta_4^3 \alpha_5 \beta^2 \delta_4^2 
\ea
\ee
The sections $\delta_4=0$ and $\delta_5=0$ give rise to new divisors $E_4$ and $E_5$ respectively.
The sections are now
\begin{equation}\begin{array}{c|c}
\text{Section} & \text{Bundle} \\ \hline
y_5 & {\cal{O}}(\sigma +3c_1 - E_1 - E_2 - E_3-E_4-E_5) \\
\zeta_4& {\cal{O}}(E_1-E_2-E_3- E_4) \\
\alpha_5 & {\cal{O}}(E_2-E_3-E_5) \\
\delta_4 & {\cal{O}}(E_4) \\
\delta_5 & {\cal{O}}(E_5) \\
w & {\cal{O}}(\sigma) \\
x_2 & {\cal{O}}(\sigma+2c_1-E_1-E_2) \\
z_1 & {\cal{O}}(S_2-E_1) \\
\beta & {\cal{O}}(E_3)
\end{array}
\end{equation}

The resolved fourfold has class
\be
{[\tilde{Y}_4]} = 6 c_1+3 \sigma -2 E_1-2 E_2-2 E_3-E_4-E_5 \,.
\ee

%%%%%%%%%%%%%%%%

\subsubsection{Constraints}

The blowups give rise to the following projective coordinates
\be
\ba
{[w, x, y]} &= [w,x_2 \zeta_4 \alpha _5^2 \beta ^3 \delta _4 \delta_5^2,y_5 \zeta_4 \alpha_5^2 \beta^4 \delta_4^2 \delta_5^3]\cr
{[x_1, y_1, z_1]}& = [x_2 \alpha _5 \beta  \delta _5,y_5 \alpha _5\beta ^2 \delta _4 \delta_5^2,z_1] \cr
[x_2, y_2,\zeta_2] &= [x_2,y_5 \beta  \delta _4 \delta_5, \zeta _4 \beta \delta_4] \cr
[y_3, \zeta_3, \alpha_3] &= [y_5 \delta _4 \delta_5, \zeta _4 \delta _4,\alpha_5 \delta_5] \cr
[y_4, \zeta_4] &= [y_5 \delta_5, \zeta_4] \cr
[y_5, \alpha_5] 
\ea
\ee
None of these combinations are allowed to simultaneously vanish.
%%%%%%%%%%%%%%%%

\subsection{Cartan Divisors}

The section $z=0$, where the $D_5$ singularity is located, splits after the blowups as
\be 
z=z_1\zeta _4 \alpha _5 \beta ^2 \delta_4 \delta_5 =0\,.
\ee
Note that the component $\delta_4=0$ is reducible, with one component given by $\zeta_4=0$.

The Cartan divisors are these six factors restricted to the resolved 4-fold $\tilde{Y_4}$, and are given by
\begin{equation}\begin{array}{c|c|c}
\text{Cartan Divisor} & \text{Component} & \text{Class in $Y_4$} \\ \hline
{\cal{D}}_{-\alpha_0}  &\left. \left(z_1=0 \right) \right| _{Y_4} & S_2-E_1 \\
{\cal{D}}_{-\alpha_1}  &\left. \left(\delta _4=0 \right) \right| _{Y_4,\zeta_4\neq0} & -E_1+E_2+E_3+2E_4 \\
{\cal{D}}_{-\alpha_2}  &\left. \left(\zeta_4 =0 \right) \right| _{Y_4} & E_1-E_2-E_3-E_4 \\
{\cal{D}}_{-\alpha_3} &\left. \left(\beta =0 \right) \right| _{Y_4} & E_3  \\
{\cal{D}}_{-\alpha_4} &\left. \left(\delta_5 =0 \right) \right| _{Y_4} & E_5\\
{\cal{D}}_{-\alpha_5}  &  \left. \left(\alpha_5=0 \right) \right| _{Y_4} & (E_2-E_3-E_5) 
\end{array}
\end{equation}

The intersection of the Cartan divisors $\{  {\cal{D}}_{-\alpha_1}, {\cal{D}}_{-\alpha_2}, {\cal{D}}_{-\alpha_3}, {\cal{D}}_{-\alpha_4}, {\cal{D}}_{-\alpha_5}, {\cal{D}}_{-\alpha_0}\}$ with the dual curves are
\be
\left(
\begin{array}{cccccc}
 -2 & 0 & 1 & 0 & 0 & 0 \\
 0 & -2 & 1 & 0 & 0 & 0 \\
 1 & 1 & -2 & 1 & 0 & 0 \\
 0 & 0 & 1 & -2 & 1 & 1 \\
 0 & 0 & 0 & 1 & -2 & 0 \\
 0 & 0 & 0 & 1 & 0 & -2
\end{array}
\right)
\,.
\ee

%%%%%%%%%%%%%%%%

\section{Matter and Yukawas}

The discriminant of the $SO(10)$ singularity has an expansion
\be\ba
\Delta = & - 16 z^7 b_2^3 b_3^2 \cr
& +\left(-27 b_3^4-8 b_1^2 b_2^2 b_3^2+72 b_2 b_4 b_3^2+4 b_1 b_2
   \left(9 b_3^2+4 b_2 b_4\right) b_3+16 b_2^2 \left(b_4^2-4 b_2 b_6\right)\right)
   z^8+O\left(z^9\right) \,.
\ea\ee
From this and the Tate algorithm table, we expect to get an enhancement to $D_6$ at $b_3=0$, corresponding to matter in the $\bold{10}$ representation, and an enhancement to $E_6$ along $b_2=0$, which corresponds to matter in the $\bold{16}$. Where both of these matter curves intersect, i.e. at $b_2=b_3=0$, we expect an enhancement to $E_7$, and therefore the Yukawa interaction point. 

In summary, we have the following enhancements:
\be
\ba
D_6: \qquad & b_3=0 \cr
E_6: \qquad & b_2=0 \cr
E_7: \qquad & b_2= b_3=0 \cr
D_7: \qquad & b_3=b_4 ^2 -4b_2 b_6=0 \,.
\ea
\ee

%%%%%%%%%%%%%%%%

\subsection{16 matter}

We expect to get matter in the $\bold{16}$ of $SO(10)$ along $z=b_2=0$.

To see the relevant root splitting, we look at one specific component of $z=0$, namely $\beta=0$. 
We can see that $\beta=b_2=0$ gives
\be 
y_5 \delta_4 (y_5 \delta_5 +b_3 \zeta_4)=0 \, ,
\ee
so it reduces to three components:
\be 
\left[\beta \right] \cdot \left[ b_2 \right] = \left[ \beta \right] \cdot \left[ y_5 \right] + \left[ \beta \right] \cdot ([ \delta_4] - [\zeta_4 ]) + [\beta] \cdot ( [b_2]- [y_5]-[\delta_4]+[\zeta_4])
\ee
The second component is specifically $ \left[ \beta \right] \cdot ([ \delta_4] - [\zeta_4 ])$ since the first Cartan divisor restricted to $b_2$ gives $\beta=0$. So $z=0$ splits into 7 components along $b_2=0$, these are
\begin{equation}\begin{array}{c|c|c}
\text{Component of} \, \left. \left( z=b_2=0 \right) \right|_{\tilde{Y}_4} & \text{Equations in} \, \tilde{Y}_4 &\text{Cartan charges} \\ \hline
\left(S_2 -E_1 \right) \cdot \left( 2c_1 - S_2 \right) & z_1 =0 & \left(0, 0, 0, 1, 0 \right) \\
\, & b_2 =0 & \, \\ \hline
\left(- E_1 + E_2 + E_3 + 2E_4 \right) \cdot \left( 2c_1 -S_2 \right) & \left. \delta_4 =0 \right|_{\zeta_4 \neq 0} & \left( -2,1,0,0,0 \right) \\
\, & b_2 =0 & \, \\ \hline
\left( E_1 -E_2 -E_3 -E_4 \right) \cdot \left( 2c_1 - S_2 \right) & \zeta_4 =0 & \left( 1, -2, 1, 0, 0 \right) \\
\, & b_2 =0 & \, \\ \hline
\left( E_3 \right) \cdot \left(  \sigma + 3 c_1 -  E_1-  E_2 -E_3-E_4 - E_5 \right) & \beta=0 & \left(1, 0, -1, 1, 0 \right) \\
\, & y_5=0 & \, \\ \hline
\left( E_3 \right) \cdot \left( -\sigma -c_1-S_2 +2E_1-E_4+E_5  \right) & \beta=0 & \left(1, 0, -1, 0, 1 \right) \\
\, & \left. b_2=0 \right|_{y_5,\delta_4 \neq 0} &\, \\ \hline
\left( E_5 \right) \cdot \left( 2c_1 - S_2 \right) & \delta_5 =0 &  \left(0, 0, 1, -2, 0 \right)\\
\, & b_2 =0 & \, \\ \hline
\left( E_2 -E_3 - E_5 \right) \cdot \left( 2c_1 - S_2 \right) & \alpha_5 =0  & \left( 0, 0, 1, 0, -2 \right) \\
\, & b_2 = 0 & \, \\ \hline

\end{array}\end{equation}

The splitting of the weight associated to the third root is
\be
-\alpha_{3}  =  ( 0,1, -2, 1, 1)\qquad \rightarrow \qquad (-2,1,0,0,0) + ( 1, 0, -1, 1, 0) + (1, 0, -1, 0, 1) \,,
\ee
with the latter two components corresponding to $-(\mu_{\bf 16} - \alpha_1 -\alpha_2 - \alpha_3 -\alpha_4-\alpha_5)$ and 
$ \mu_{\bf 16}  - \alpha_2 -2\alpha_3 -\alpha_4 -\alpha_5 $, which confirms the matter in the $\mathbf{16}$ representation.

%%%%%%%%%%%%%
\subsection{10 Matter}

We expect to get matter in the $\mathbf{10}$ along $z=b_3=0$. 

To see the relevant root splitting for the matter in the vector representation, we look at another specific component of $z=0$, namely $\delta_5=0$.
 We can see that the Cartan divisor $\delta_5=0$ splits here to
\be 
\alpha_5 (b_2 x_2^2 +b_4 x_2 \beta + b_6 \beta^2)
\ee
where we have set equal to 1 any variables which cannot vanish simultaneously with $\delta_5$. So this has split into three components, the first is just another Cartan divisor restricted to $b_3=0$. The expression in brackets is factorised if we assume $b_2\neq0$. We do this as $b_2=b_3=0$ corresponds to a Yukawa coupling which we consider in the next section. We call the 2 factors $\gamma_+$ and $\gamma_-$ and they have the same homology class. Overall, $\delta_5=0$ reduces as
\be 
[\delta_5] \cdot [b_3] = [\delta_5] \cdot [\alpha_5] + 2 \times \frac{[\delta_5]\cdot([b_2]-[\alpha_5])}{2}
\ee

So we see that $z=b_3 =0$ splits into 7 components, as one would expect from a "$D_6$" enhancement. The seven components are
\begin{equation}\begin{array}{c|c|c}
\text{Component of} \, \left. \left( z=b_3=0 \right) \right|_{\tilde{Y}_4} & \text{Equations in} \, \tilde{Y}_4 &\text{Cartan charges} \\ \hline
\left(S_2 -E_1 \right) \cdot \left( 3c_1 - 2S_2 \right) & z_1 =0 & \left(0, 0, 0, 1, 0 \right) \\
\, & b_2 =0 & \, \\ \hline
\left(- E_1 + E_2 + E_3 + 2E_4 \right) \cdot \left( 3c_1 -2S_2 \right) & \left. \delta_4 =0 \right|_{\zeta_4 \neq 0} & \left( -2,1,0,0,0 \right) \\
\, & b_2 =0 & \, \\ \hline
\left( E_1 -E_2 -E_3 -E_4 \right) \cdot \left( 3c_1 - 2S_2 \right) & \zeta_4 =0 & \left( 1, -2, 1, 0, 0 \right) \\
\, & b_2 =0 & \, \\ \hline
\left( E_3 \right) \cdot \left(  3c_1-2S_2 \right) & \beta=0 & \left(0, 1, -2, 1, 1 \right) \\
\, & b_3=0 & \, \\ \hline
\frac{1}{2}\left( E_5 \right) \cdot \left(  3c_1-2S_2-E_2+E_3+E_5 \right) & \delta_5=0 & \left(0, 0, 0, -1, 1 \right) \\
\, & \gamma_+=0 &\, \\ \hline
\frac{1}{2}\left( E_5 \right) \cdot \left(3c_1-2S_2-E_2+E_3+E_5 \right) & \delta_5 =0 &  \left(0, 0, 0, -1, 1 \right)\\
\, & \gamma_- =0 & \, \\ \hline
\left( E_2 -E_3 - E_5 \right) \cdot \left( 3c_1 - 2S_2 \right) & \alpha_5 =0  & \left( 0, 0, 1, 0, -2 \right) \\
\, & b_2 = 0 & \, \\ \hline
\end{array}\end{equation}

The $\delta_5=0$ root splits as
\be 
\left(0,0,1,-2,0 \right) \qquad \rightarrow \qquad \left(0,0,1,0,-2 \right) + \left(0,0,0,-1,1 \right) + \left(0,0,0,-1,1 \right) \,.
\ee
The first component is a Cartan divisor, but the other two are both given by
\be 
\mu_{\bf{10}} -\alpha_1 -\alpha_2 -\alpha_3 -\alpha_4 \,, 
\ee
So indeed this corresponds to matter in the $\bf{10}$.

%%%%%%%%%%%%%%%%%

\subsection{Yukawa Coupling}

We expect to get a Yukawa interaction at the point which corresponds to an "$E_7$" enhancement which is given by $b_2=b_3=0$. We could think of this as a further enhancement of the "$E_6$" curve, therefore instead of looking at how the six components of $z=0$ split, we study how the seven components of $z=b_2=0$ split.

Firstly we can see from above that the third component of $\beta=b_2=0$ will split further to
\be 
y_5 \delta_5=0 \,.
\ee 
Also, we have $\delta_5=b_2=b_3=0$ gives
\be 
\alpha_5 \beta (b_4 x_2 +b_6 \beta)=0
\ee 
We already had the first two components, but the last is new.
So altogether we see that $z=b_2=b_3=0$ has 7 components, given by
\begin{equation}\begin{array}{c|c|c}
\text{Component of} \, \left. \left( z=b_2=b_3=0 \right) \right|_{\tilde{Y}_4} & \text{Equations in} \, \tilde{Y}_4 &\text{Cartan charges} \\ \hline
\left(S_2 -E_1 \right) \cdot \left( 2c_1 - S_2 \right) \cdot \left(3c_1-2S_2 \right) & z_1 =0 & \left(0, 0, 0, 1, 0 \right) \\
\, & b_2 =0 & \, \\
\, &b_3=0 &\, \\ \hline
\left(- E_1 + E_2 + E_3 + 2E_4 \right) \cdot \left( 2c_1 -S_2 \right)\cdot \left(3c_1-2S_2 \right) & \left. \delta_4 =0 \right|_{\zeta_4 \neq 0} & \left( -2,1,0,0,0 \right) \\
\, & b_2 =0 & \, \\ 
\, & b_3=0 & \, \\\hline
\left( E_1 -E_2 -E_3 -E_4 \right) \cdot \left( 2c_1 - S_2 \right) \cdot \left(3c_1-2S_2 \right)& \zeta_4 =0 & \left( 1, -2, 1, 0, 0 \right) \\
\, & b_2 =0 & \, \\
\, & b_3=0 & \, \\ \hline
\left( E_3 \right) \cdot \left(  \sigma + 3 c_1 -  E_1-  E_2 -E_3-E_4 - E_5 \right) \cdot & \beta=0 & \left(1, 0, -1, 1, 0 \right) \\
 \left(3c_1-2S_2 \right) & y_5=0 & \, \\
 \, & b_3=0 & \, \\ \hline
\left( E_3 \right) \cdot \left( -\sigma -c_1-S_2 +2E_1-E_4+E_5  \right) \cdot  & \beta=0 & \left(0, 0, 0, -1, 1 \right) \\
 \left(3c_1-2S_2 \right) & \left. b_2=0 \right|_{y_5,\delta_4 \neq 0} &\, \\ 
 \, & \delta_5=0 & \, \\ \hline
\left( E_5 \right) \cdot \left( 2c_1 - S_2 \right) \cdot \left(3c_1-2S_2 \right)& \delta_5 =0 &  \left(0, 0, 0, -1, 1 \right)\\
\, & b_2 =0 & \, \\
\, & \left. b_3=0 \right|_{\alpha_5,\beta\neq 0} \, \\  \hline
\left( E_2 -E_3 - E_5 \right) \cdot \left( 2c_1 - S_2 \right) & \alpha_5 =0  & \left( 0, 0, 1, 0, -2 \right) \\
\, & b_2 = 0 & \, \\ \hline

\end{array}\end{equation}

We can see that this corresponds to the Yukawa coupling $\mathbf{16} \times \mathbf{16} \times \mathbf{10} $ by approaching this point along the $\mathbf{16}$ matter curve,where we can see the splitting:

\be 
\ba 
(1,0,-1,0,1) \qquad \rightarrow &  \qquad (1,0,-1,1,0) + (0,0,0,-1,1) \cr
(\mu_{16}  -\alpha_2 -2\alpha_3 -\alpha_4 -\alpha_5) \qquad  \rightarrow & \qquad -(\mu_{16} -\alpha_1 -\alpha_2 -\alpha_3 -\alpha_4 -\alpha_5)  \cr
& \qquad + (\mu_{10} -\alpha_1 -\alpha_2 -\alpha_3 -\alpha_4)
\ea
\ee

This gives the desired Yukawa coupling.
It is interesting to note here that our "$E_7$" enhancement has only 7 components instead of the expected 8, this is similar to what was shown to happen with $E_6$ in \cite{ey} (see also \cite{mt}), where one node was missing.

\subsection{D7 enhancement}

We expect to get a "D7" enhancement at $b_3=b_4^2-4b_2 b_6=0$. Starting from the "D6" enhancement, i.e. with $b_3=0$, the effect of setting $b_4^2-4b_2 b_6=0$ is to make $\gamma_+=\gamma_- \equiv \gamma$, so these two previously separate components are now the same.  
   So despite it being a "D7" enhancement, we actually only get six components instead of the expected eight. These are:
  \begin{equation}\begin{array}{c|c|c}
\text{Component of} \, \left. \left( z=b_3=b_4^2-4b_2b_6=0 \right) \right|_{\tilde{Y}_4} & \text{Equations in} \, \tilde{Y}_4 &\text{Cartan charges} \\ \hline
\left(S_2 -E_1 \right) \cdot \left( 3c_1 - 2S_2 \right) \cdot \left(8c_1-6S_2 \right) & z_1 =0 & \left(0, 0, 0, 1, 0 \right) \\
\, & b_2 =0 & \, \\
\, & b_4^2-4b_2b_6=0 & \, \\ \hline
\left(- E_1 + E_2 + E_3 + 2E_4 \right) \cdot \left( 3c_1 -2S_2 \right) \cdot \left(8c_1-6S_2 \right)& \left. \delta_4 =0 \right|_{\zeta_4 \neq 0} & \left( -2,1,0,0,0 \right) \\
\, & b_2 =0 & \, \\ 
\, & b_4^2-4b_2b_6=0 & \, \\\hline
\left( E_1 -E_2 -E_3 -E_4 \right) \cdot \left( 3c_1 - 2S_2 \right) \cdot \left(8c_1-6S_2 \right)& \zeta_4 =0 & \left( 1, -2, 1, 0, 0 \right) \\
\, & b_2 =0 & \, \\ 
\, & b_4^2-4b_2b_6=0 & \, \\ \hline
\left( E_3 \right) \cdot \left(  3c_1-2S_2 \right)\cdot \left(8c_1-6S_2 \right) & \beta=0 & \left(0, 1, -2, 1, 1 \right) \\
\, & b_3=0 & \, \\
\, & b_4^2-4b_2b_6=0 & \, \\ \hline
\frac{1}{2}\left( E_5 \right) \cdot \left(3c_1-2S_2-E_2+E_3+E_5 \right) \cdot  (8c_1-6S_2)& \delta_5 =0 &  \left(0, 0, 0, -1, 1 \right)\\
\, & \gamma =0 & \, \\
\, & b_4^2-4b_2b_6=0 & \, \\ \hline
\left( E_2 -E_3 - E_5 \right) \cdot \left( 3c_1 - 2S_2 \right) \cdot \left(8c_1-6S_2 \right)& \alpha_5 =0  & \left( 0, 0, 1, 0, -2 \right) \\
\, & b_2 = 0 & \, \\
\, & b_4^2-4b_2b_6=0 & \, \\ \hline
\end{array}\end{equation}
Here, at the point of enhancement, we see that the two previously separate $\mathbf{10}$ matter curves become one, we believe that this corresponds to a $\mathbf{10} \times \mathbf{10} \times \mathbf{1}$ coupling, we do not see a curve for the singlet as it is not part of the GUT divisor.

\section{G-Flux}
The existence of the (2,2) form G-flux is a requirement of the heterotic - F-theory duality. For local geometries, the G-flux can be constructed from 
Heterotic string data in terms of spectral covers \cite{tw3,dw2}. For global geometries, \cite{sm4,sm5} proposed a global approach to G-fluxes which was successfully
applied to the case of $A_4$ singularities in \cite{sm1} \footnote{For other recent approaches to G-flux see \cite{other}}.

In our work, we are firstly computing the local flux and then the global one.

\subsection{Local G-Flux}
\subsubsection{Tate divisor in $Y_4$}
Our original 4-fold $Y_4$ is given by
\be 
y^2 w + b_1 z x y w + 
 b_3 z^2 y w^2  =  x^3 + b_2 z x^2 w + b_4 z^3 x w^2 + b_6 z^5 w^3 ,
 \ee
and we are interested in the Tate divisor
\be 
wz \left( b_2 x^2 + b_4 z^2 xw + b_6 z^4 w^2 -b_1 xy -b_3 zyw \right) \, .
\ee
Along the Tate divisor we have $wy^2=x^3$. By assuming $w\neq0$, we can then rewrite the Tate divisor in terms of the $t=y/x$ as
\be 
z \left(b_2 t^4 +b_4 z^2 t^2 +b_6 z^4 -b_1 t^5 -b_3 z t^3 \right)
\ee 
By setting $s=z/t$ and holding $s$ fixed in the limit $t \rightarrow 0$, $z \rightarrow 0$, we obtain 
\be 
st^5 \left( b_2  +b_4  s^2 + b_6 s^4 - b_3 s \right) 
\ee 
\subsubsection{Tate divisor in resolved $\tilde{Y}_4$}

Now we look at the resolved Calabi-Yau $\tilde{Y}_4$, setting
\be 
\ba 
x & = x_2  \zeta_4 \alpha_5^2 \beta^3 \delta_4 \delta_5^2 \cr
y & = y_5  \zeta_4 \alpha_5^2 \beta^4 \delta_4^2 \delta_5^3 \cr
z &= z_1 \zeta_4 \alpha_5 \beta^2 \delta_4 \delta_5 \, .
\ea 
\ee 

This makes
\be 
\ba 
t & = \frac{y}{x}=\frac{y_5 \beta \delta_4 \delta_5}{x_2} \cr
s & = \frac{z}{t}=\frac{z_1 x_2 \zeta_4 \alpha_5 \beta}{y_5}
\ea 
\ee
The limit $t \rightarrow 0$ with $s$ held fixed can be achieved by taking the limit $\delta_4 \rightarrow 0$ or $\delta_5 \rightarrow 0$.

Now we take the total transform of the Tate divisor,
\be 
wy^2-x^3=0
\ee 
which gives
\be 
\zeta_4^2 \alpha_5^4 \beta^8 \delta_4^3  \delta_5^6 \left(wy_5^2\delta_4-x_2^3 \zeta_4 \alpha_5^2 \beta \right) =0
\ee 

The term
\be 
wy_5^2\delta_4-x_2^3 \zeta_4 \alpha_5^2 \beta =0
\ee
is the proper transform of the Tate divisor, which we then restrict to the resolved $\tilde{Y}_4$.

With this restriction, the proper transform of the Tate divisor is reducible, with components given by $\zeta_4=0$, $z_1=0$, and the remainder. 

To see this, set $\zeta_4=0$. By using (\ref{NonvanishingY4pairs}), we see that we cannot have $w=0$, $y_5=0$, $\alpha_5=0$ or $\delta_5=0$, so we set these equal to one. The Tate divisor equation is now
\be 
\delta_4=0 \,,
\ee
and the equation for the resolved $\tilde{Y}_4$ also becomes
\be 
\delta_4=0 \,.
\ee 
Which means that the Tate divisor equation is automatically satisfied.

For the component $z_1=0$, we cannot have $\alpha_5$, $\beta$ or $\delta_5$  equal to zero, so again we set these equal to one. The Tate divisor equation takes the form
\be 
wy_5^2\delta_4-x_2^3 \zeta_4=0 \, ,
\ee
and the equation for $\tilde{Y}_4$
\be 
wy_5^2 \delta_4=x_2^3 \zeta_4 \,.
\ee
This again means that the Tate divisor equation is satisfied automatically in $\tilde{Y}_4$. So we define the Tate divisor by
\be 
\mathcal{C}_{\text{Tate}}=[wy_5^2\delta_4-x_2^3 \zeta_4 \alpha_5^2 \beta]\cdot[\tilde{Y}_4]-[\zeta_4=0]\cdot[\tilde{Y}_4]-[z_1=0]\cdot[\tilde{Y}_4]
\ee
Which is in the class
\be 
3\sigma+6c_1-S_2-2E_1-E_2-E_3-2E_5 \, .
\ee
Its intersection with the Cartan divisors takes the form:
\be 
\mathcal{C}_{\text{Tate}}\cdot_{\tilde{Y}_4}\Sigma_{\alpha_i}=(0,0,0,1,0) \times 4 \, .
\ee

\subsubsection{Local limit}
As discussed before, the local limit may be either $\delta_4 \rightarrow 0$ or $\delta_5 \rightarrow 0$. However, if we set $\delta_4$ equal to zero in the Tate divisor equation, and take $x_2=\alpha_5=1$ as they cannot vanish together with $\delta_4$, then we obtain
\be 
\zeta_4 \beta=0 ,.
\ee 
But since $\zeta_4=0$ is one of the reducible components we remove from the Tate divisor, this leaves $\beta=0$, which means that $s=z/t$ becomes zero, instead of being held fixed. So the local limit here must be given by $\delta_5 \rightarrow 0$, so we intersect with $E_5$:
\be \ba 
\mathcal{C}_{\text{Tate}} \cdot_{\tilde{Y}_4}E_5 &= [wy_4^2-x_2^3\alpha_4^2\beta\zeta_3]\cdot_{\tilde{Y}_4}E_5-[\zeta_3=0]\cdot_{\tilde{Y}_4}E_5 -[z_1=0]\cdot_{\tilde{Y}_4}E_5\cr
&= [wy_4^2-x_2^3\alpha_4^2\beta\zeta_3]\cdot_{\tilde{Y}_4}E_5
\ea 
\ee 
where the last terms vanish because we cannot have $\delta_5=z_1=0$ or $\delta_5=\zeta_4=0$.
Setting $\delta_5=0$, and $w=z_1=\zeta_4=1$, the equation for the Tate divisor becomes

\be 
 y_5^2=x_2^3 \alpha_5^2 \beta \,.
\ee
Notice that if $\alpha_5=0$, then $y_5=0$, but these two conditions are not allowed to hold simultaneously, so we set $\alpha_5=1$.
This implies that the equation for $\tilde{Y}_4$ becomes
\be 
b_3  y_5 =  b_2 x_2^2  \zeta_4 \alpha_ +b_4  x_2   \beta +b_6  \beta^2,.
\ee

If we now set $x_2=0$, the Tate divisor equation gives $y_5=0$ and so the equation for  $\tilde{Y}_4$ becomes
\be 
0=b_6 \beta^2 .
\ee
For generic $b_6$ this sets $\beta=0$. Under the identification $\delta_5=0$, $\zeta_4=\delta_4=1$,
the coordinates of the $\mathbb{P}^2$ from the second blow up become
\be 
[x_2,0,\beta]
\ee
$x_2=0$ would imply $\beta=0$ so $x_2=0$ is not allowed and we can set $x_2=1$.
The Tate divisor equation is
\be 
y_5^2=\beta
\ee 
which, after substituting it into $\tilde{Y}_4$, gives the required spectral equation
\be 
b_6y_5^4 +b_4y_5^2 -b_3y_5 +b_2=0
\ee

\subsubsection{Local flux}
To construct the local flux, as in \cite{sm1}, we firstly construct the surface $\mathcal{S}_{p^*D}$ by
\be 
\mathcal{S}_{p^*D}=\mathcal{C}_{\text{Tate}}\cdot D-(3\sigma+6c_1) \cdot D
\ee 
which only intersects the Cartan root $\mathcal{D}_{-\alpha_4}$, and we have made a subtraction to get the required orthogonality properties.

We now consider the intersections between the matter surfaces of the $\mathbf{16}$ (which we can either take as $\beta=y_5=0$ or $\beta=\delta_5=0$) and 
$\mathcal{C}_{\text{Higgs,loc}}=\mathcal{C}_{\text{Tate}}\cdot_{\tilde{Y}_4}E_3$, along the curve
\be 
y_5=b_2=0
\ee
So for $\mathcal{S}_{\sigma \cdot \mathcal{C}}$ we take the surface of the $\mathbf{16}$ which implies $y_4=b_2=0$ i.e. $\beta=y_5=0$. In order for this to intersect Cartan divisors other than $\mathcal{D}_{-\alpha_4}$, we must make subtractions from it to obtain the desired intersection properties.
We end up with
\be 
\ba 
\mathcal{S}_{\sigma \cdot \mathcal{C}} &=[\beta]\cdot_{\tilde{Y}_4}[y_5]-[b_2]\cdot_{\tilde{Y}_4}(\mathcal{D}_{-\alpha_2}+2\mathcal{D}_{-\alpha_3}+\mathcal{D}_{-\alpha_4}+\mathcal{D}_{-\alpha_5})\cr 
&= E_3\cdot_{\tilde{Y}_4}(\sigma+3c_1-E_1-E_2-E_3-E_4-E_5) - (2c_1-S_2) \cdot_{\tilde{Y}_4} (E_1-E_4)
\ea 
\ee

Using these two surfaces, we can construct a local G-flux with a traceless combination
\be 
\ba 
G_{\text{local}} &=4\mathcal{S}_{\sigma \cdot \mathcal{C}}-\mathcal{S}_{p^*(2c_1-S_2)} \cr
&= c1\cdot_{\tilde{Y}_4} (-4E_1 +2E_2+14E_3+8E_4+4E_5) + S_2\cdot_{\tilde{Y}_4} (2E_1-E_2-E_3-4E_4-2E_5) \cr
&-4E_3 \cdot_{\tilde{Y}_4} (E_1+E_2+E_3+E_4+E_5) \,.
\ea 
\ee
This can be further simplified by using the relations in the appendix to
\be 
\ba 
G_{\text{local}} =&c_1\cdot_{\tilde{Y}_4}(4E_1-6E_2+6E_3-8E_4+4E_5) +S_2 \cdot_{\tilde{Y}_4}(-2E_1+3E_2-5E_3+4E_4-2E_5) \cr
& +4E_3\cdot_{\tilde{Y}_4} E_4-4E_3\cdot_{\tilde{Y}_4} E_5\,.
\ea 
\ee

The ramification divisor can be computed using 
\be 
r=\mathcal{C}_{\text{Higgs,loc}}\cdot \left(\mathcal{C}_{\text{Higgs,loc}}-\sigma-\sigma_{\infty} \right)
\ee 
whose odd part is given by
\be 
\mathcal{S}_r^{(\text{odd})}=c_1\cdot E_2 +c_1 \cdot E_3\,.
\ee 
We have the quantization condition that
\be 
\alpha G_{\text{local}}  + \frac{1}{2} \mathcal{S}_r^{(\text{odd})}
\ee
be integrally quantized, for some choice of $\alpha \in \mathbb{C}$. Looking at $G_{\text{local}}$ and $\mathcal{S}_r^{(\text{odd})}$, \cite{sak} has argued that that this can not be done generically. We can satisfy this requirement by imposing extra conditions on $B_3$ and $S_2$. Now if we look at the last two terms of G, we see that they do not depend on $B_3$ or $S_2$, and so to satisfy the quantization condition we require $\alpha \in \frac{\mathbb{Z}}{4}$. Since our condition only concerns the non-integer part, we then have four cases to consider: $\alpha \in \mathbb{Z}$, $\alpha \in \mathbb{Z}+\frac{1}{4}$, $\alpha \in \mathbb{Z} + \frac{1}{2}$ and $\alpha \in \mathbb{Z} + \frac{3}{4}$. 

Firstly, for $\alpha \in \mathbb{Z}$, we simply require that the odd part of the ramification divisor vanishes, which can be done by choosing the base such that $c_1(B_3)$ is even.

For $\alpha \in \mathbb{Z} + \frac{1}{2}$, we require $c_1(S_2)$ even, for later use we note that this means that the odd parts of $c_1(B_3)$ and $S_2$ now match.

The last two possibilities, $\alpha \in \mathbb{Z} \pm \frac{1}{4}$ both give the same condition, which is that the class $S_2$ be a multiple of 4.
\subsection{General G-Flux}
In order to compute the global G-flux, we first need to compute the 2nd Chern class, since the G-flux is quantized according to
\be 
G+\frac{1}{2}c_2(\tilde{Y}_4) \in H^4(\tilde{Y}_4,\mathbb{Z}) \, .
\ee 
We compute the second Chern class using the formula from \cite{Aluffi}:
\be 
\ba 
c(\tilde{X}_5)=&c(X_5)\frac{(1+E_1)(1+\sigma+2c_1-E_1)(1+\sigma+3c_1-E_1)(1+S_2-E_1)}{(1+\sigma+2c_1)(1+\sigma+3c_1)(1+S_2)} \cr
& \times \frac{(1+E_2)(1+\sigma+2c_1-E_1-E_2)(1+\sigma+3c_1-E_1-E_2)(1+E_1-E_2)}{(1+\sigma+2c_1-E_1)(1+\sigma+3c_1-E_1)(1+E_1)} \cr
&\times \frac{(1+E_3)(1+\sigma+3c_1-E_1-E_2-E_3)(1+E_1-E_2-E_3)(1+E_2-E_3)}{(1+\sigma+3c_1-E_1-E_2)(1+E_1-E_2)(1+E_2)} \cr
&\times \frac{(1+E_4)(1+\sigma+3c_1-E_1-E_2-E_3-E_4)(1+E_1-E_2-E_3-E_4)}{(1+\sigma+3c_1-E_1-E_2-E_3)(1+E_1-E_2-E_3)} \cr
&\times \frac{(1+E_5)(1+\sigma+3c_1-E_1-E_2-E_3-E_4-E_5)(1+E_2-E_3-E_5)}{(1+\sigma+3c_1-E_1-E_2-E_3-E_4)(1+E_2-E_3)} \, .
\ea
\ee 
We can then restrict this to $\tilde{Y}_4$ by
\be 
c(\tilde{Y}_4)=\frac{c(\tilde{X}_5)}{1+3\sigma+6c_1-2E_1-2E_2-2E_3-E_4-E_5} \,.
\ee 
After expanding this out, the odd part of $c_2(\tilde{Y}_4)$ is given by
\be 
\ba 
c_2^{(odd)}(\tilde{Y}_4)&=c_1\cdot (E_1+E_2+E_3+E_4+E_5)+S_2\cdot E_1+E_1\cdot E_2 \cr
 &+E_1 \cdot E_5 + E_2 \cdot E_3 +E_4\cdot E_4 +E_5 \cdot E_5\,.
\ea 
\ee
Simplifying using the relations in the appendix gives
\be 
c_2^{(odd)}(\tilde{Y}_4)=c_1\cdot E_2+c_1 \cdot E_3\,.
\ee

As well as the quantization condition, we also require that the G-flux be orthogonal to horizontal and vertical surfaces in $Y_4$. Also, to preserve the $SO(10)$ symmetry we require that the intersection between $G$ and the Cartan divisors vanish. As in \cite{sm1}, this restricts $G$ to be a linear combination of $c1\cdot_{\tilde{Y}_4}E_i$, $S_2\cdot_{\tilde{Y}_4}E_i$ and $E_i\cdot_{\tilde{Y}_4}E_j$. Using the relations between exceptional divisors in the appendix, we can eliminate all combinations of $E_i\cdot_{\tilde{Y}_4}E_j$ except for two of them, which we choose to be $E_3\cdot_{\tilde{Y}_4}E_4$ and $E_3\cdot_{\tilde{Y}_4}E_5$. So overall $G$ is then given by
\be 
G=\frac{1}{2} c_1\cdot(E_2+E_3) + \sum_{i=1}^{5} (a_i c_1 \cdot E_i + b_i S_2 \cdot E_i) +p E_3\cdot E_4 + q E_3 \cdot E_5 
\ee

Here $a_i$, $b_i$, $p$ and $q$ are integers, and the first two terms are present to enforce the quantization condition.

However, using this expression for $G$ and requiring it to vanish when intersected with the Cartan divisors does not give integer answers for all of $a_i$, $b_i$, $p$ and $q$. This was of course to be expected, since in the previous section we saw that $G$ cannot be quantized without imposing extra conditions. We now solve for the G-flux for two of these conditions.

1) For $c_1(B_3)$ even, the odd part of $c_2$ vanishes and our general form for $G$ is 
\be 
G= \sum_{i=1}^{5} (a_i c_1 \cdot E_i + b_i S_2 \cdot E_i) +p E_3\cdot E_4 + q E_3 \cdot E_5 
\ee \,.
The requirement that this vanish when intersected with the Cartan divisors gives the one parameter solution
\be 
\ba 
a_1 &= 4n \cr
a_2 &= -6n \cr
a_3 &= 6n \cr
a_4 &= -8n \cr
a_5 &= 4n \cr
b_1 &= -2n \cr
b_2 &= 3n \cr
b_3 &= -5n \cr
b_4 &= 4n \cr
b_5 &= -2n \cr
p &= 4n \cr
q &= -4n \,,
\ea 
\ee
where $n$ is an integer. So the flux is
\be 
\ba 
G=&n \left( c_1 \cdot \left( 4E_1-6E_2+6E_3-8E_4+4E_5 \right) +S_2 \cdot \left( -2E_1 +3E_2 -5E_3 +4E_4 -2E_5 \right) \right. \cr & \left. +4 E_3 \cdot E_4 -4 E_3 \cdot E_5 \right)
\ea
\ee

2) For $c_1(S_2)$ to be even, the odd part of $c_2$ can now be written as $S_2 \cdot E_2+S_2 \cdot E_3$, and so we take $G$ to be 
\be 
G=\frac{1}{2} S_2\cdot(E_2+E_3) + \sum_{i=1}^{5} (a_i c_1 \cdot E_i + b_i S_2 \cdot E_i) +p E_3\cdot E_4 + q E_3 \cdot E_5 \,.
\ee
As before, we impose that $G$ does not intersect any of the Cartan divisors, and obtain the one parameter solution
\be 
\ba 
a_1 &= 2+4n \cr
a_2 &= -3-6n \cr
a_3 &= 3+6n \cr
a_4 &= -4-8n \cr
a_5 &= 2+4n \cr
b_1 &= -1-2n \cr
b_2 &= 1+3n \cr
b_3 &= -3-5n \cr
b_4 &= 2+4n \cr
b_5 &=-1-2n \cr
p &=2+4n \cr
q &=-2-4n \,,
\ea 
\ee
again with $n$ integral. This gives
\be 
\ba 
G= &\left(n+\frac{1}{2} \right) \left( c_1 \cdot \left( 4E_1 -6E_2 +6E_3 -8E_4 +4E_5 \right) + S_2 \left( -2E_1 +3E-2 -5E_3 +4E_4-2E_5 \right) \right. \cr
& \left. + 4E_3 \cdot E_4 -4E_3 \cdot E_5 \right)
\ea 
\ee
3) When the class of $S_2$ a multiple of 4, one also gets the same answer as in the local case. So altogether we see that with each of the three possible conditions, the global and local fluxes match.

We can now intersect the flux with the matter surfaces. we take our matter surfaces as
\be 
\ba 
\mathcal{S}_{\mathbf{16}} & = E_3 \cdot \left( \sigma +3c_1 -E_1 -E-2 -E_3 - E_4 -E_5 \right) \cr
\mathcal{S}_{\mathbf{10}} & = \frac{1}{2} E_5 \cdot \left( 3c_1 -2S_2 -E_2 +E_3 +E_5 \right) \,.
\ea 
\ee
 We then obtain
 \be 
 \ba 
 G \cdot_{\tilde{Y}_4} \mathcal{S}_{\mathbf{16}} &= \alpha  (6c_1-5S_2) \cdot_{S_2} ( 2c_1-S_2) \cr 
 G \cdot_{\tilde{Y}_4} \mathcal{S}_{\mathbf{10}} &=0 \,.
 \ea
 \ee
 Both of these are in agreement with \cite{so101}, the zero chirality of the $\mathbf{10}$ is of course a problem, and is resolved there by taking a factorised spectral cover. We assume that this would also work here, but leave the computations for future work.

\section*{Acknowledgements}

We would like to thank Sakura Sch\"{a}fer-Nameki for her very kind help, assistance and encouragement during the development of this project. We also thank Moritz K\"{u}ntzler for helpful discussions.

%%%%%%%%%%%%%%%%%%%%%%%%
%%%%%%%%%%%%%%%%%%%%%%%%

\appendix
\section{Intersection Relations}

\subsection{Intersection Relations in $X_5$}
Here we list relations that hold in $X_5$ coming from the constraints we get at each blowup concerning the non-vanishing of sets of homogeneous coordinates.

\be
\ba
0 &= \sigma \left(\sigma + 2 c_1 \right) \left( \sigma +3c_1 \right) \cr
0 &= \left(\sigma + 2 c_1 -E_1 \right) \left( \sigma +3c_1 -E_1\right) \left(S_2 - E_1\right) \cr
0 &= \left(\sigma + 2 c_1 - E_1 -E_2 \right) \left( \sigma +3c_1 -E_1 - E_2\right) \left(E_1 - E_2\right) \cr
0 &= \left(\sigma +3c_1 - E_1 - E_2 - E_3 \right) \left(E_1 - E_2 - E_3 \right) \left( E_2 - E_3 \right) \cr
0 &= \left(\sigma +3c_1 - E_1 - E_2 - E_3 -E_4 \right) \left( E_1 - E_2 - E_3 - E_4 \right) \cr
0 &= \left(\sigma+3c_1-E_1-E_2-E_3-E_4 -E_5 \right) \left(  E_2-E_3 - E_5 \right)
\ea
\ee
As all blowups are done in the $w=1$ patch, we have that
\be 
0=\sigma \cdot E_i
\ee
since the class $\sigma$ given by $w=0$ cannot then intersect any of the exceptional divisors.
\subsection{Intersection Relations in $\tilde{Y}_4$}
Consider the sets of coordinates associated to each blow up and see which ones cannot simultaneously vanish:
\be
\ba
{[w, x, y]} &= [w,x_2 \zeta_4 \alpha _5^2 \beta ^3 \delta _4 \delta_5^2,y_5 \zeta_4 \alpha_5^2 \beta^4 \delta_4^2 \delta_5^3]\cr
{[x_1, y_1, z_1]}& = [x_2 \alpha _5 \beta  \delta _5,y_5 \alpha _5\beta ^2 \delta _4 \delta_5^2,z_1] \cr
[x_2, y_2,\zeta_2] &= [x_2,y_5 \beta  \delta _4 \delta_5, \zeta _4 \beta \delta_4] \cr
[y_3, \zeta_3, \alpha_3] &= [y_5 \delta _4 \delta_5, \zeta _4 \delta _4,\alpha_5 \delta_5] \cr
[y_4, \zeta_4] &= [y_5 \delta_5, \zeta_4] \cr
[y_5, \alpha_5] 
\ea
\ee

With these relations and the equation for $\tilde{Y}_4$, we see that we cannot have solutions to any of the following equations:
\begin{equation}
\ba
x_2=\zeta_4 &=0 \cr
x_2=\beta &=0 \cr
x_2=\delta_4 &=0 \cr
y_5=z_1 &=0 \cr
y_5=\zeta_4 &=0 \cr
y_5=\alpha_5 &=0 \cr
z_1=\alpha_5 &=0 \cr
z_1 =\beta &=0 \cr
z_1 =\delta_5 &=0 \cr
\zeta_4=\alpha_5 &=0 \cr
\zeta_4=\delta_5 &=0 \cr
\alpha_5=\delta_4 &=0 \cr
\delta_4=\delta_5 &=0 \, .
\ea
\label{NonvanishingY4pairs}
\end{equation}

These imply
\be 
\ba 
(\sigma +2c_1 -E_1-E_2) \cdot_{\tilde{Y}_4} (E_1-E_2-E_3-E_4) &=0 \cr
(\sigma+2c_1-E_1-E_2) \cdot_{\tilde{Y}_4} (E_3) &=0 \cr
(\sigma+2c_1-E_1-E_2) \cdot_{\tilde{Y}_4} (E_4) &=0 \cr
(\sigma+3c_1-E_1-E_2-E_3-E_4-E_5) \cdot_{\tilde{Y}_4} (S_2-E_1) &=0 \cr
(\sigma+3c_1-E_1-E_2-E_3-E_4-E_5) \cdot_{\tilde{Y}_4} (E_1-E_2-E_3-E_4) &=0 \cr
(\sigma+3c_1-E_1-E_2-E_3-E_4-E_5) \cdot_{\tilde{Y}_4} (E_2-E_3-E_5) &=0 \cr
(S_2-E_1) \cdot_{\tilde{Y}_4} (E_2-E_3-E_5) &=0 \cr
(S_2-E_1) \cdot_{\tilde{Y}_4} (E_3) &=0 \cr
(S_2-E_1) \cdot_{\tilde{Y}_4} (E_5) &=0 \cr
(E_1-E_2-E_3-E_4) \cdot_{\tilde{Y}_4} (E_2-E_3-E_5) &=0 \cr
(E_1-E_2-E_3-E_4) \cdot_{\tilde{Y}_4} (E_5) &=0 \cr
(E_2-E_3-E_5) \cdot_{\tilde{Y}_4} (E_4) &=0 \cr
(E_4) \cdot_{\tilde{Y}_4} (E_5) &=0 \, .
\ea
\ee

There are two more relations we can get by considering the surface $z_1=0$ in $\tilde{Y}_4$. Since $z_1=0$ means that we cannot have $y_5$ $\alpha_5$, $\beta$ or $\delta_5$ vanishing, we set these equal to 1, this leaves
\be 
w \delta_4=x_2^3 \zeta_4 \,.
\ee
By setting $\zeta_4=0$ implies $\delta_4=0$, and vice versa, so these are equivalent, giving
\be 
(S_2-E_1)\cdot_{\tilde{Y}_4} (E_1-E_2-E_3-E_4) = (S_2-E_1) \cdot_{\tilde{Y}_4} (E_4) \,,
\ee
or
\be 
(S_2-E_1)\cdot_{\tilde{Y}_4} (E_1-E_2-E_3-2E_4) = 0\,.
\ee
Also we see that setting $w=0$ implies $x_2=0$ and vice versa, so these are also equivalent, which gives the relation
\be 
(S_2-E_1) \cdot_{\tilde{Y}_4} (2c_1-E_1-E_2) =0 \,.
\ee

\section{$SO \left( 10 \right) $ weights and roots}
\begin{equation}\begin{array}{c|c}
\hline
\text{Cartan charges of $\bold{10}$} & \text{Root} \\ \hline
\left( 1,0,0,0,0 \right) & \mu_{10} \\
\left( -1,1,0,0,0 \right) & \mu_{10} - \alpha_1 \\
\left( 0,-1,1,0,0 \right) & \mu_{10} -\alpha_1 - \alpha_2 \\
\left( 0,0,-1,1,1 \right) & \mu_{10} -\alpha_1 - \alpha_2 -\alpha_3 \\
\left( 0,0,0,-1,1 \right) & \mu_{10} -\alpha_1 -\alpha_2 -\alpha_3 -\alpha_4 \\
\left( 0,0,0,1,-1 \right) & \mu_{10} -\alpha_1 -\alpha_2 -\alpha_3 -\alpha_5 \\
\left( 0,0,1,-1,-1 \right) & \mu_{10} -\alpha_1 -\alpha_2 -\alpha_3 -\alpha_4 -\alpha_5 \\
\left( 0,1,-1,0,0 \right) & \mu_{10} -\alpha_1 -\alpha_2 -2\alpha_3 -\alpha_4 -\alpha_5 \\
\left( 1,-1,0,0,0 \right) & \mu_{10} -\alpha_1 -2\alpha_2 -2\alpha_3 -\alpha_4 -\alpha_5 \\
\left( -1,0,0,0,0 \right) & \mu_{10} -2\alpha_1 -2\alpha_2 -2\alpha_3 -\alpha_4 -\alpha_5 \\ \hline
\text{ Cartan charges of $\bold{16}$} & \text{Root} \\ \hline
\left( 0,0,0,0,1 \right) & \mu_{16} \\
\left( 0,0,1,0,-1 \right) & \mu_{16} -\alpha_5 \\
\left( 0,1,-1,1,0 \right) & \mu_{16} -\alpha_3 -\alpha_5 \\
\left( 1,-1,0,1,0 \right) & \mu_{16} -\alpha_2 -\alpha_3 -\alpha_5 \\
\left( 0,1,0,-1,0 \right) & \mu_{16} -\alpha_3 -\alpha_4 -\alpha_5 \\
\left( -1,0,0,1,0 \right) & \mu_{16} -\alpha_1 -\alpha_2 -\alpha_3 -\alpha_4 \\
\left( 1,-1,1,-1,0 \right) & \mu_{16} -\alpha_2 -\alpha_3 -\alpha_4 -\alpha_5 \\
\left( -1,0,1,-1,0 \right) & \mu_{16} -\alpha_1 -\alpha_2 -\alpha_3 -\alpha_4 -\alpha_5 \\
\left( 1,0,-1,0,1 \right) & \mu_{16}  -\alpha_2 -2\alpha_3 -\alpha_4 -\alpha_5 \\
\left( -1,1,-1,0,1 \right) & \mu_{16} -\alpha_1 -\alpha_2 -2\alpha_3 -\alpha_4 -\alpha_5 \\
\left( 1,0,0,0,-1 \right) & \mu_{16} -\alpha_2 -2\alpha_3 -\alpha_4 -2\alpha_5 \\
\left( 0,-1,0,0,1 \right) & \mu_{16} -\alpha_1 -2\alpha_2 -2\alpha_3 -\alpha_4 -\alpha_5 \\
\left( -1,1,0,0,-1 \right) & \mu_{16} -\alpha_1 -\alpha_2 -2\alpha_3 -\alpha_4 -2\alpha_5 \\
\left( 0,-1,1,0,-1 \right) & \mu_{16} -\alpha_1 -2\alpha_2 -2\alpha_3 -\alpha_4 -2\alpha_5 \\
\left( 0,0,-1,1,0 \right) & \mu_{16} -\alpha_1 -2\alpha_2 -3\alpha_3 -\alpha_4 -2\alpha_5 \\
\left( 0,0,0,-1,0 \right) & \mu_{16} -\alpha_1 -2\alpha_2 -3\alpha_3 -2\alpha_4 -2\alpha_5
\end{array}
\end{equation}

\end{document}